**Title:** Estimation of the attributable fraction for time to event outcomes using an inverse probability of exposure weighted Kaplan-Meier estimator

**Authors:** Denis Talbot[1,2], Miceline Mésidor[1,2], Kossi Clément Trenou[1,2], Mathilde Lavigne-Robichaud[1,2], Xavier Trudel[1,2], Aida Eslami[1,3].

**Affiliations:**

[1]Département de médecine sociale et préventive, Université Laval, Québec, Canada

[2]Centre de recherche du CHU de Québec – Université Laval Québec, Canada

[3]Centre de recherche de l'Institut universitaire de cardiologie et de pneumologie de Québec – Université Laval Québec, Canada

**Corresponding author:** Denis Talbot, Département de médecine sociale et préventive,

Université Laval, Québec, Canada

E-mail: denis.talbot@fmed.ulaval.ca

**Abstract**

Population attributable fractions aim to quantify the proportion of the cases of an outcome (for example, a disease) that would have been avoided had no individuals in the population been exposed to a given exposure. This quantity thus plays a crucial role in epidemiology and public health, notably to guide policies, interventions or to assess the burden of a disease due to a particular exposure. Various statistical methods have been proposed to estimate attributable fractions using observational data. When time-to-event data are used, several of these formulas yield invalid results. Alternative valid formulas are available but remain scarcely used. We propose a new estimator of the attributable fraction that is both conceptually simple and easy to implement using common statistical software. Our proposed estimator makes use of the Kaplan-Meier estimator to address censoring and potentially non-proportional hazards, as well as inverse probability weighting to control confounding. Nonparametric bootstrap is proposed to produce inferences. A simulation study is used to illustrate and compare our proposed estimator to several alternatives. The results showcase the bias of many commonly used traditional approaches and the validity of our estimator under its working assumptions.

**Introduction**

Widely used in public health and epidemiology, the population attributable fraction (AF) aims to assess the impact of an exposure on a disease (1, 2). As it is a causal measure, it can be defined in a counterfactual way and corresponds to the ratio between the reduction in the incidence of the outcome that could be achieved if the population had been entirely unexposed (i.e., counterfactual) and the observed incidence (i.e., factual) (3).

The AF has been developed for a binary outcome and is often estimated using the two following formula $AF = \frac{(p_d(RR-1))}{RR}$, where $p_d$ is the prevalence of the exposure among the cases and RR the risk ratio between the exposed and non-exposed individuals and $AF = \frac{pe(RR-1)}{pe(RR-1)+1}$, where $p_e$ is the prevalence of the total population exposed (4). As mentioned by several authors, the second formula is valid only when there is no confounding or effect modification (5, 6). Censored time-to-event data are frequent in epidemiology. Some studies have used the traditional formulas above by replacing the risk ratio (RR) by the hazard ratio (HR) when using such data (7-9). However, for a time-to-event variable, the traditional formula is biased not only because it does not allow for censoring but also because it is sensitive to collider bias (10). Several alternative methods adapted to time-to-event data have been proposed in the literature (2, 11-14). However, these methods remain scarcely used. This may partly be due to the mathematical complexity of some of these methods, which make them not accessible to epidemiologists. Moreover, some of these methods are difficult to use because of the unavailability of software implementing them.

The objective of this paper is to propose an approach based on the Kaplan-Meier estimator weighted by the inverse of the probability of treatment (or exposure) to adjust for confounding factors and calculate AFs with a time-to-event outcome. To simplify the implementation of this method, R code is provided on GitHub (https://github.com/detal9/IPTW_AF). The paper is structured as follows. We first present our proposed estimator. Then we use simulated data to compare our proposal to alternatives for which code is readily available in a variety of scenarios. We conclude with a short discussion on the strengths and limitations of the various approaches as well as recommendations.

**Method**

*Proposed approach: Inverse probability of exposure weighted Kaplan-Meier*

Using a counterfactual notation, the AF at time *t* for a time-to-event outcome can be defined as:

$$AF_t = 1 - \left(\frac{1 - S_t^0}{1 - S_t}\right),$$

where $S_t^0$ represents the cumulative incidence of the outcome at time *t* that would be observed if everyone was unexposed, and $S_t$ represents the observed cumulative incidence of the outcome (2). Note that with a time-to-event outcome, the AF needs to be calculated at a given time *t*, which emphasizes that the proportion of the cases that are attributable to a given exposure can vary in time.

Because of censoring, neither $S_t^0$, nor $S_t$ can generally be validly estimated as a simple proportion of subjects who did not have the outcome at time *t*. However, under a noninformative censoring assumption, $S_t$ can be consistently estimated using an approach that is familiar to most epidemiologists, the Kaplan-Meier estimator (15), meaning that the estimates converge to the true value as sample size increases. Estimating $S_t^0$ not only requires accounting for censoring, but also typically requires adjustment for confounding factors. Inverse probability of exposure weighting (IPW) is becoming an increasingly popular approach to adjust for confounding. The general approach consists in weighting each subject by the inverse of the probability of receiving the exposure level they actually received, conditional on their potential confounder values (16). In the weighted sample, exposed and unexposed subjects are expected to have similar characteristics, thus mimicking a randomized trial relative to the measured confounders. We propose to estimate $S_t^0$ using the Kaplan-Meier estimator fitted among unexposed subjects, weighted according to the inverse probability of exposure weights. This estimator is consistent under the no unmeasured confounding assumption and the noninformative censoring assumptions (17).

In summary, the AF at time *t* can be estimated as follows:

1. Estimate $S_t$ by the estimated "survival" (no outcome) probability at time *t* using the Kaplan-Meier estimator in the full sample.
2. Fit a model for *P(A = 0|L)*, where *A* is the exposure and *L* the measured potential confounders, in the full sample. This can be a logistic regression model if *A* is binary or a multinomial model if *A* has more than two levels, for example.
3. For each unexposed individual, compute the predicted value of the previous model and then compute the inverse probability of exposure weight $w = 1/\hat{P}(A = 0|L)$.
4. Estimate $S_t^0$ by estimated survival probability at time *t* using the Kaplan-Meier estimator among unexposed subjects only, weighting each subject by the inverse probability of exposure weight *w*.
5. Substitute the estimated $S_t$ and $S_t^0$ in the AF formula.

Inferences (e.g., confidence intervals) can be produced by using the nonparametric bootstrap. Because the Kaplan-Meier estimator can be expressed as an m-estimator (18), a sandwich estimator of the variance of the proposed AF estimator can also be obtained (19).

A key strength of the proposed estimator is its relative conceptual and implementation simplicity. In addition, it is adapted to censored time-to-event outcomes, unlike traditional formulas. Finally, it does not require making a proportional hazards assumption, unlike approaches based on the

Cox model (e.g., (14)). However, the noninformative censoring assumption can be unreasonable in some situations. This assumption entails that knowing the censoring time of an individual should provide no information on their time-to-event. If there are factors that affects both the time-to-event and the time to censoring, this assumption can be violated. For example, informative censoring would be present if those with a greater risk of the outcome also tend to be more quickly lost to follow-up (censored). A common situation where noninformative censoring would be reasonable to assume is when censoring only occurs because of the administrative end of follow-up. In which case, censoring times are fixed by design and thus independent of the time-to-event.

In the next section, we present a simulation study aimed at evaluating and comparing our proposed estimator to several alternatives in various scenarios.

*Simulation scenarios*

We considered five different simulation scenarios, inspired by an ongoing project. This project concerns the estimation of the proportion of cases of coronary heart disease attributable to psychosocial stressor at work exposure among white-collar workers. The parameters of the simulation were chosen such that the sample size, the proportion of exposed individuals, the number of events and the exposure hazard ratio were approximately that of the men in these data.

In each scenario, we generated two independent baseline covariates. The covariate $L_1$ was generated as a uniform covariate over the interval [20, 50], inspired by the age of participants at baseline. The second covariate $L_2$ was generated as binary variable with probability 0.5.

In Scenario 1, the data were generated such that there was no confounding, no censoring and that the proportional hazard assumption would be met. The exposure $A$ was generated as a binary variable with probability *expit(-1)*, where *expit(x) = exp(x)/[1 + exp(x)]* is the inverse of the logit function. The time to the event was generated according to a Cox model with baseline hazard following a Weibull distribution with parameters $\lambda = 0.00003$ and *a = 2.1*, and regression coefficients $\exp(0.5A + 0.1L_1 + L_2)$. The exposure and the time to event were generated in the same way in Scenarios 2-4.

There was also no censoring in Scenario 2 and the proportional hazard assumption was again met, but confounding was present. In this Scenario, the exposure was generated as a binary variable with probability $expit(-5 + 0.1L_1 + L_2)$. The exposure was also generated in this way in Scenarios 2-5.

The data in Scenario 3 was generated such that the proportional hazard assumption was met, and that random (noninformative) censoring was present. Censoring was generated as a log-normal distribution with mean on the log scale *log(5)* and standard deviation on the log scale of *log(1.2)*. Censoring was also generated in this way in Scenario 5.

The proportional hazard assumption was also met in Scenario 4, but time to censoring was generated as a function of covariates and exposure. More precisely, the time to censoring was generated according to a Cox with baseline hazard following a Weibull distribution with parameters $\lambda = 0.00008$ and *a =2.4*, and regression coefficients $\exp(0.5A + 0.1L_1 + L_2)$.

In Scenario 5, the data were generated such that the proportional hazard assumption would not be verified, but that censoring was random, as in Scenario 3. The time to event was generated in two parts, both generated according to different Cox models with the same baseline hazard as previously described. However, the regression parameters differed between both parts. In the first part, the regression parameters were $\exp(1.4 + 0.1L_1)$ whereas they were $\exp(3 + A + 0.1L_1 + L_2)$. The total time to event was equal to the minimum between four and the time generated from the first part. The time generated according to the second part was then added if the time from the first part was equal to four. The exposure was generated as $\exp(-2 - 0.1L_1 + L_2)$.

For each scenario, we generated 1000 samples of size *n = 3000*. The top part of table 1 summarizes the characteristics of the five simulation scenarios.

*Simulation analysis*

The true 5-year AF was estimated by Monte Carlo simulation, by generating the counterfactual time to event under no exposure and the observed time-to-event using a large dataset of size *n = 5,000,000* to minimize the Monte Carlo error. The true AFs were 13.3% in Scenario 1, 20.8% in Scenarios 2-4, and 10.1% in Scenario 5.

In each dataset, we estimated the AF using the observed data with seven different estimators. We chose these estimators because they were simple to implement, either because their formula was simple or because of the availability of R code.

Two estimators were based on the formula:

$$AF = \frac{pd(RR-1)}{RR},$$

where the RR was substituted either by the exposure hazard ratio of a Cox proportional hazard adjusted for $L_1$ and $L_2$ as covariates, or by the exposure hazard ratio of an IPW-adjusted Cox proportional hazard model. We designate these estimators as *pd.cov* and *pd.ipw*, respectively. A 95% confidence interval was obtained by substituting the bounds of the 95% confidence interval for the HR into the preceding formula, where a conservative 95% confidence interval for the IPW-adjusted HR was obtained using a robust estimator. Similarly, two estimators were based on the formula:

$$AF = \frac{pe(RR-1)}{pe(RR-1)+1},$$

where the RR was again substituted by either the covariate-adjusted or the IPT-adjusted exposure hazard ratio of a Cox model. We call these estimators as *pe.cov* and *pe.ipw*, respectively. As mentioned in the introduction, the four previous estimators are not expected to be appropriate with a time-to-event outcome.

We also estimated the AF using the *getAF* function from the *averisk* package in R and call the estimator *getAF*. This approach has been developed for binary outcomes and calculates the mean of the AF for a set of exposure variables adjusted for confounding (20). This function has been previously used with a time-to-event outcome (21), despite modeling the outcome using a logistic regression model instead of using a time-to-event model. For this approach, we used a binary indicator that the event was observed within five years. A 95% confidence interval was obtained using 200 bootstrap replicates.

The sixth and seventh estimators are the *AFcoxph* function from the *AF* package in R, and our proposed inverse probability of exposure weighted Kaplan-Meier estimator with 200 bootstrap replicates, respectively. In the following, we call these estimators *AFcoxph* and *IPW.KM*, respectively. Briefly, the *AFcoxph* function incorporates confounder-adjusted estimation of the AF through Cox regression and thus assumes proportional hazards (14). However, this estimator allows for the censoring time and the time-to-event to depend on common factors, as long as those factors are included in the Cox model.

In each scenario, we calculated the Monte Carlo bias of each estimator as the difference between the average of the estimates and the true 5-year AF. We also computed the standard deviation (SD) of the estimates produced by each estimator and the proportion of the replicates in which the 95% confidence interval included the true AF.

*Results*

The results of the simulation study are presented in the bottom part of Table 1. The four traditional formulas yield biased estimates and 95% confidence intervals that include the true AF less often then expected in all scenarios. The *getAF* produced estimators with relatively low bias in Scenarios 1-3 but with substantial bias in Scenarios 4-5. However, the 95% confidence intervals included the true effect much more often than expected, which means that the statistical uncertainty is overestimated. The *AFcoxph* produced unbiased estimates with appropriate coverage of 95% confidence intervals in all Scenarios except Scenario 5 where the proportional hazard assumption wasn't verified. Our proposed estimator, *IPW.KM*, similarly produced good results in all Scenarios except the one where censoring was affected by covariates. When comparing unbiased estimators together, the *AFcoxph* estimator had lower standard deviations than either *getAF* and *IPW.KM*, which had relatively similar standard deviations.

Table 1: Summary of the simulation scenarios and results of the simulation study

|  | Scenario 1 | | | Scenario 2 | | | Scenario 3 | | | Scenario 4 | | | Scenario 5 | | |
| --- | --- | --- | --- | --- | --- | --- | --- | --- | --- | --- | --- | --- | --- | --- | --- |
| *Censoring* | No | | | No | | | Yes, noninformative | | | Yes, informative | | | Yes, noninformative | | |
| *Confounding* | No | | | Yes | | | Yes | | | Yes | | | Yes | | |
| *Proportional hazards* | Yes | | | Yes | | | Yes | | | Yes | | | No | | |
| *Estimator* | Bias | SD | Cover | Bias | SD | Cover | Bias | SD | Cover | Bias | SD | Cover | Bias | SD | Cover |
| *pd.cov* | 1.1 | 1.7 | 64.6 | 2.0 | 2.2 | 69.4 | 2.0 | 5.7 | 86.9 | -0.6 | 4.5 | 86.0 | 0.8 | 2.2 | 81.7 |
| *pd.ipw* | -3.3 | 1.3 | 19.8 | -4.9 | 2.6 | 53.2 | 1.0 | 5.8 | 91.1 | -6.4 | 5.3 | 66.9 | -0.5 | 2.3 | 85.7 |
| *pe.cov* | 1.5 | 1.4 | 78.6 | -4.4 | 1.5 | 20.3 | -4.3 | 4.6 | 83.4 | -4.5 | 3.8 | 77.6 | 7.7 | 3.4 | 38.2 |
| *pe.ipw* | -4.1 | 1.0 | 7.5 | -10.4 | 1.9 | 0.0 | -5.2 | 4.6 | 80.1 | -10.2 | 4.2 | 36.3 | 4.9 | 3.5 | 72.7 |
| *getAF* | 0.0 | 3.8 | 99.9 | -0.5 | 5.3 | 99.1 | -0.2 | 5.8 | 99.5 | -6.6 | 6.7 | 96.7 | -2.8 | 2.7 | 99.5 |
| *AFcoxph* | 0.1 | 1.3 | 95.1 | 0.0 | 1.8 | 96.0 | -0.1 | 5.3 | 94.4 | -0.2 | 4.4 | 95.2 | 1.4 | 2.2 | 90.4 |
| *IPW.KM* | -0.1 | 3.9 | 94.6 | -0.1 | 5.7 | 93.6 | -0.1 | 6.4 | 93.4 | -5.8 | 6.6 | 84.3 | -0.2 | 3.1 | 93.4 |

**Discussion**

Although it has been mentioned in multiple publications that substituting hazard ratios in place of risk ratios in traditional AF formula is invalid, this practice remains prevalent. While several alternatives that are adapted to time-to-event data have been introduced, they remain scarcely used. We hypothesized that the mathematical complexity or the lack of available software were, at least partly, responsible for the gap between theory and analytical practice. As such, we proposed an inverse probability weighted Kaplan-Meier approach to estimate the AF, which is both conceptually simple to understand and relatively easy to implement using common software. We further provide R code on GitHub implementing our approach (https://github.com/detal9/IPTW_AF).

We also employed a simulation study to evaluate our approach and compare it with alternatives. Importantly, this simulation study highlights that traditional formulas can induce substantial bias when hazard ratios are substituted in place of risk ratios, even in ideal scenarios where there is no confounding and no censoring. The formula based on the overall prevalence of exposure (*pe*) had particularly high bias, which may be unsurprising considering this formula should not be used when confounding is present (5, 6). The formula based on computing the average of the adjusted AFs (*getAF*) was also observed to be biased in our simulation study (20). Again, this is an unsurprising result since this method is not adapted to censored time-to-event data. Our proposed estimator performed well in scenarios with noninformative censoring, but poorly when informative censoring was present. Opposingly, the estimator based on a Cox regression model performed well when censoring and time-to-event depended on common measured factors, but poorly under non-proportional hazards. Both the Cox regression and the weighted Kaplan-Meier approaches are relatively simple conceptually and can easily be implemented using available code in the R software. When both approaches are valid (i.e., with noninformative censoring and proportional hazards), the estimator based on the Cox regression seems preferable since it has lower variability. In the ongoing analysis that motivated the simulation study, the proportional hazards assumption appeared to be violated, whereas noninformative was at least approximately reasonable since most censoring were due to the administrative end of the follow-up.

In conclusion, we join others in warning against using traditional AF formulas with time-to-event outcomes. Using such formulas can induce serious biases.